\documentclass[twoside,10pt]{article}
\usepackage{1}

\usepackage{hyperref}
\usepackage{graphicx}
\usepackage{multirow}

\title{\Large \bf $\ensuremath{t\bar{t}}$ charge asymmetry at hadron colliders}
\author{A. Chapelain}
\date{}
\begin{document}
\maketitle

\begin{center}
\vspace*{-0.3cm}
{\it  Irfu/SPP, CEA-Saclay, France\\
e-mail: antoine.chapelain@cea.fr }
\end{center}

\vspace{0.3cm}

\begin{center}
{\bf Abstract}\\
\medskip
\parbox[t]{10cm}{\footnotesize
We present the current status for the measurements of the $\ensuremath{t\bar{t}}$ charge asymmetry at the Tevatron and LHC colliders. }
\end{center}

\section{Introduction} \label{sec:intro}
At NLO, QCD predicts the top quark to be emitted preferentially in the direction of the incoming quark,
while the top antiquark in the direction of the incoming antiquark.
This charge asymmetry comes mainly from the interference between 
$\ensuremath{q\bar{q}}~\rightarrow~\ensuremath{t\bar{t}}$ tree diagram with the NLO box diagram,
and from the interference of initial and final state radiations ($\ensuremath{q\bar{q}} \rightarrow \ensuremath{t\bar{t}} g$).
Results from the CDF and D0~\cite{CDF_51_ttbar, D0_54_ttbar} collaborations have driven a
lot of attention because some of the measured asymmetries were significantly higher than the SM predictions.
In this note we present the most recent results for the CDF, D0, ATLAS and CMS collaborations and compared them to the latest prediction based on the Standard Model~\cite{BS}.

\section{Observables}

The Tevatron and LHC colliders present different initial state particles at different energies. The Tevatron is a proton-antiproton collider at
$\sqrt{s}=1.96$~TeV and the LHC is a proton-proton collider at $\sqrt{s}=7~\text{and}~8$ TeV. At the Tevatron $\ensuremath{t\bar{t}}$ pairs
are mainly produced through quark-antiquark anihilation and there the laboratory frame is equivalent to the partonic
rest frame. Thus at the Tevatron the charge asymmetry results into a forward-backward asymmetry (FB).
At the LHC $\ensuremath{t\bar{t}}$ pairs are mainly produced through gluon-gluon fusion which does not contribute to the charge asymmetry.
In the subdominant $\ensuremath{q\bar{q}} \rightarrow \ensuremath{t\bar{t}}$ process, the antiquark originates from the proton sea leading the $\ensuremath{t\bar{t}}$ system to be boosted in the direction of the top quark.
Indeed the incoming quark carries on average a higher momentum than the incoming antiquark. In that case the charge asymmetry results into a forward-central asymmetry (FC). 

We therefore define two different observables at the Tevatron and at the LHC to measure the $\ensuremath{t\bar{t}}$ charge asymmetry:
\begin{eqnarray}
\text{Tevatron} & A^{\ensuremath{t\bar{t}}}_{{FB}} & =  \frac{N(\Delta y > 0) - N(\Delta y < 0)}{N(\Delta y > 0) + N(\Delta y < 0)}, \\
& \text{where} & \Delta y  = y_t - y_{\bar{t}}, \nonumber \\ 
\text{LHC} & A_{C} & = \frac{N(\Delta |y| > 0) - N(\Delta |y| < 0)}{N(\Delta |y| > 0) + N(\Delta |y| < 0)}, \\
& \text{where} & \Delta |y| = |y_t| - |y_{\bar{t}}|, \nonumber 
\label{eq:01}
\end{eqnarray}
$y_t$ and $y_{\bar{t}}$ are top quark and antiquark rapidity, respectively.

We also measure the $\ensuremath{t\bar{t}}$ charge asymmetry based on the leptons coming from the decay of the $W$ boson coming from the top quark decay:
\begin{eqnarray}
\text{Tevatron} & \ensuremath{A^{\ell\ell}} & =  \frac{N(\Delta\eta > 0) - N(\Delta\eta < 0)}{N(\Delta\eta > 0) + N(\Delta\eta < 0)}, \\
& \ensuremath{A^{\ell}_{\rm FB}} & = \frac{N(q \times \eta>0) - N(q \times \eta<0)}{N(q\times \eta>0) + N(q \times \eta<0)}, \\
& \text{where} & \Delta\eta  = \eta_{\ell^+} - y_{\ell^-}, \nonumber \\ 
\text{LHC} & A^{\ell\ell}_C & = \frac{N(\Delta|\eta| > 0) - N(\Delta|\eta| < 0)}{N(\Delta|\eta| > 0) + N(\Delta|\eta| < 0)}, \\
& \text{where} & \Delta|\eta| = |\eta_{\ell^+}| - |\eta_{\ell^-}|, \nonumber 
\label{eq:01}
\end{eqnarray}
$\eta_{\ell^+}$ and $y_{\ell^-}$ are positive and negative lepton pseudorapidity, respectively.

The interest of measuring such an asymmetry is that we do not need to reconstruct the $\ensuremath{t\bar{t}}$ kinematic and the lepton kinematic is
well measured. The leptonic asymmetry is also sensible to the top quark polarization if any. 

\section{Top quark signature and reconstruction}

The top quark decays almost $100\%$ of the time into a $W$ boson and in a $b$ quark. We therefore classify the $\ensuremath{t\bar{t}}$ final state according to the $W$ boson
decay mode. 

The $\ell$+jets channel, where one $W$ boson decays hadronically and the other leptonically,
is characterized by one isolated lepton, at least four jets and missing energy due to the presence of a neutrino escaping the detector.
This channel has a good production rate ($\sim 45\%$ of all $\ensuremath{t\bar{t}}$ events) and a reasonable amount of background. The main backgrounds in this channel are
$W$+jets events estimated using simulation and data, and multijet production where one jet mimics a lepton. The later is estimated using data.
The dilepton channel, where both $W$ decay leptonically, is characterized by two oppositely charged leptons, at least two jets
and missing energy due to the two neutrinos.
This channel suffers from a smaller production rate ($\sim 5\%$) but has little background. The main backgrounds come from Drell-Yan process estimated using the 
simulation and $W$+jets and multijets background mimicking leptons estimated in data.

Beside the above requirement, the event selection, which aims at increasing the $\ensuremath{t\bar{t}}$ fraction in the analyzed sample, uses topological criteria as well as well as $b$-quark identification.

For the \ensuremath{t\bar{t}}-based asymmetry, the reconstruction of the \ensuremath{t\bar{t}} kinematics is needed.
This reconstruction is performed using kinematic fitters.
In the reconstruction algorithm, the different lepton-jet permutations, the experimental resolutions, the $b$-quark identification are taken into account.
The mass of the $W$ boson and the top quark are fixed to their world average values within their widths. In the dilepton channel we additionally need to make some assumption about the neutrinos
kinematic since their presence leaves the system unconstrained.

\section{Measurement}

After reconstruction the \ensuremath{t\bar{t}} kinematic, we can measure the raw asymmetry, i.e. the asymmetry observed in the detector.
To do so we need to subtract the estimated background from data.
At this level we cannot compare the measurements between experiments (CDF and D0 on the one hand and ATLAS and CMS on the other and) due to different detector effects and different acceptance cuts.
The raw distribution need then to be unfolded to correct for these effects and get back to the production level asymmetry.

\section{Results}

\begin{figure}[htt]
\begin{center}
\includegraphics[width=5cm]{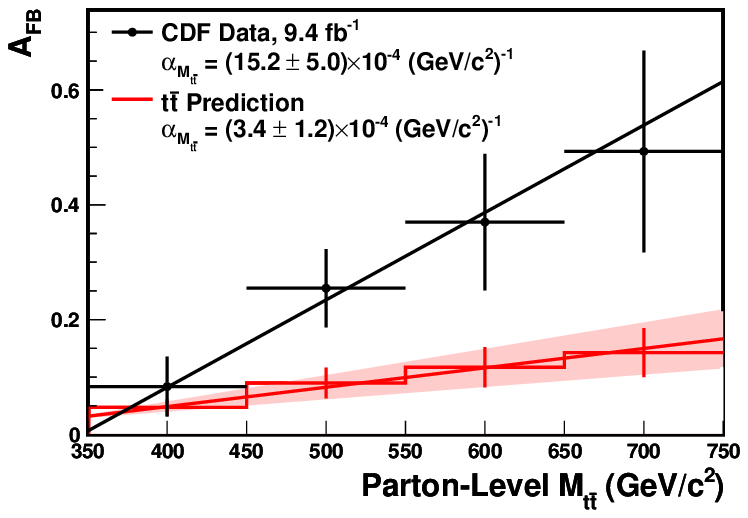}
\includegraphics[width=5cm]{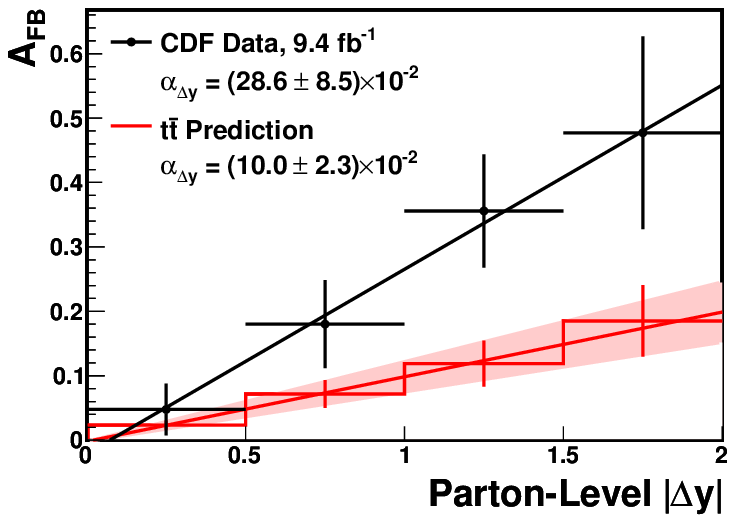}
\end{center}
\vspace{-0.5cm} \caption{$\ensuremath{A^{t\bar{t}}_{\rm FB}}$ as a function the invariant mass and rapidity of the \ensuremath{t\bar{t}} system in CDF~\cite{CDF_94_ttbar}.} 
\label{fig:1}
\end{figure}

Table~\ref{tab:tev} shows the inclusive production \ensuremath{t\bar{t}}-based and lepton-based asymmetries measured at the Tevatron compared to the predictions.
We observe that there are differences up to about two standard deviations (SD) between measurement and predictions.
Figure~\ref{fig:1} shows the differential measurement of the $\ensuremath{t\bar{t}}$ asymmetry as a function of the invariant mass and the rapidity of the \ensuremath{t\bar{t}} system performed by CDF~\cite{CDF_94_ttbar}.
We observe a significant difference between measurements and prediction up to about three SD.
D0 does not observe such a difference.

Table~\ref{tab:lhc} presents the inclusive production \ensuremath{t\bar{t}}-based and lepton-based asymmetries measured at the LHC compared to the predictions.
The measurements are consistent with the predictions. 
ATLAS and CMS performed differential measurements of the \ensuremath{t\bar{t}} asymmetry as a function the invariant mass, the transverse momentum and the rapidity of the \ensuremath{t\bar{t}}.
We observe consistency between the measurements and the predictions also for for highly boosted \ensuremath{t\bar{t}} system.

The final Tevatron measurements as well as the new LHC measurements performed at 8~TeV are expected to be published soon.

\begin{table}[hbt] \noindent\caption{Inclusive production \ensuremath{t\bar{t}}-based and lepton-based asymmetry measured at the Tevatron. All values are given in \%.}\vskip3mm\tabcolsep5pt
\begin{center}
\renewcommand{\arraystretch}{1.3}{
{\footnotesize\begin{tabular}{l | r c l | r c l r c l }
\hline\hline
& \multicolumn{3}{c|}{$\ensuremath{A^{t\bar{t}}_{\rm FB}}$} & \multicolumn{3}{c}{$\ensuremath{A^{\ell}_{\rm FB}}$} & \multicolumn{3}{c}{$\ensuremath{A^{\ell\ell}}$} \\
\hline
CDF $\ell$+jets  & $16.4$ & $\pm$ & $4.5$~\cite{CDF_94_ttbar} & $9.4$ & $\pm$ & $^{3.2}_{2.9}$\cite{CDF_94_lep} & & $-$ & \\
CDF dilepton & & $-$ & &$7.2$ & $\pm$ & $6.0$\cite{CDF_91_lep} & $7.6$ & $\pm$ & $8.1$\cite{CDF_91_lep} \\
D0 $\ell$+jets & $19.6$ & $\pm$ & $6.5$~\cite{D0_54_ttbar} & $4.7$ & $\pm$ & $^{2.6}_{2.7}$\cite{D0_97_lep} & & $-$ & \\
D0 dilepton & & $-$ & & $4.4$ & $\pm$ & $3.9$\cite{D0_97_dilep} & $12.3$ & $\pm$ & $5.6$\cite{D0_97_dilep}  \\
\hline
Prediction~\cite{BS} & $8.8$ & $\pm$ & $0.6$  & $3.8$ & $\pm$ & $0.3$ & $4.8$ & $\pm$ & $0.4$\\
\hline\hline
\end{tabular}}}
\end{center}
\label{tab:tev}
\end{table}

\begin{table}[hbt] \noindent\caption{Inclusive production \ensuremath{t\bar{t}}-based and lepton-based asymmetry measured at the LHC.
All values are given in \%.}\vskip3mm\tabcolsep5pt
\begin{center}
\renewcommand{\arraystretch}{1.3}{
{\footnotesize\begin{tabular}{l | r c l | r c l}
\hline\hline
& \multicolumn{3}{c}{$A^{\ell\ell}$} &  \multicolumn{3}{c}{$A^{\ell\ell}_C$} \\
\hline
ATLAS $\ell$+jets & $0.6$ & $\pm$ & $1.0$~\cite{ATLAS_5_ljets} & & $-$ &  \\
ATLAS dilepton &  $5.7$ & $\pm$ & $2.8$~\cite{ATLAS_5_dil} & $2.3$ & $\pm$ & $1.4$~\cite{ATLAS_5_dil} \\
CMS $\ell$+jets  & $0.5$ & $\pm$ & $0.9$~\cite{CMS_20_ljets} & & $-$ &  \\
CMS dilepton & $5.0$ & $\pm$& $^{4.4}_{5.8}$\cite{CMS_5_dil} & $1.0$ & $\pm$ & $1.6$~\cite{CMS_5_dil} \\
\hline
Prediction~\cite{BS} & $1.23$ & $\pm$ & $0.05$ & $0.70$ & $\pm$ & $0.03$ \\
\hline\hline
\end{tabular}}}
\end{center}
\label{tab:lhc}
\end{table}

\end{document}